\documentclass[twoside,11pt]{article}
\usepackage{a4wide,cite,latexsym,amsfonts,amssymb,exscale,epsfig,bbm}
\usepackage[centertags,sumlimits,intlimits,namelimits,reqno]{amsmath}

\usepackage{amsthm}
\theoremstyle{definition}

\newtheorem{theorem}{Theorem}[section]
\newtheorem{lemma}[theorem]{Lemma}

\newtheorem{proposition}[theorem]{Proposition}
\newtheorem{remark}[theorem]{Remark}
\newtheorem{definition}[theorem]{Definition}

\def\draft#1{}%
 
\renewcommand{\theequation}{\thesection.\arabic{equation}}%
\makeatletter
\@addtoreset{equation}{section}
\makeatother

\newcounter{mathletter}%
\newcommand{\bmathletter}{%
  \refstepcounter{equation}%
  \setcounter{mathletter}{\value{equation}}%
  \setcounter{equation}{0}%
  \renewcommand{\theequation}{%
    \mbox{\thesection.\arabic{mathletter}\alph{equation}}}}%
\newcommand{\emathletter}{\setcounter{equation}{\value{mathletter}}}%
\newenvironment{mathletters}{\bmathletter}{\emathletter}%

\newenvironment{myenumerate}{%
  \begin{enumerate}
  \setlength{\itemsep}{0pt}
  \setlength{\parskip}{0pt}}{\end{enumerate}}

\def\openone{\mathbbm{1}}%
\def\acknowledgements{\section*{Acknowledgements}}%

\def\dopreprint{\hfill{\small\thepreprint}\\}%
\def\preprint#1{\def\thepreprint{#1}}%

\def\ket#1{\left|#1\right>}

\def\sym#1{{\mathcal #1}}
\def\emph#1{{\sl #1\/}}

\let\phi=\varphi
\let\theta=\vartheta
\let\epsilon=\varepsilon

\let\hat=\widehat
\let\tilde=\widetilde

\def\id{\mathop{\rm id}\nolimits}

\def\id{\mathop{\rm id}\nolimits}

\def\min{\mathop{\rm min}\nolimits}
\def\qdet{\mathop{\rm qdet}\nolimits}

\def\address#1{\date{{\sl #1}\\\ \\\theversion}\gdef\date##1{}}%
\def\version#1{\gdef\theversion{#1}}%

\makeatletter
\newfont{\@aidxte}{cmsy10}
\newfont{\@aidxel}{cmsy10 scaled 1095}
\newfont{\@aidxtw}{cmsy10 scaled 1200}
\newlength\@aidxtexvi
\newlength\@aidxtexvii
\newlength\@aidxelxvi
\newlength\@aidxelxvii
\newlength\@aidxtwxvi
\newlength\@aidxtwxvii
\newcommand{\alignidx}[1]{%
  \@aidxtexvi=\fontdimen16\@aidxte
  \@aidxtexvii=\fontdimen17\@aidxte
  \@aidxelxvi=\fontdimen16\@aidxel
  \@aidxelxvii=\fontdimen17\@aidxel
  \@aidxtwxvi=\fontdimen16\@aidxtw
  \@aidxtwxvii=\fontdimen17\@aidxtw
    {\mbox{$%
    \fontdimen16\@aidxte=2.9pt
    \fontdimen17\@aidxte=2.9pt
    \fontdimen16\@aidxel=3.1pt
    \fontdimen17\@aidxel=3.1pt
    \fontdimen16\@aidxtw=3.3pt
    \fontdimen17\@aidxtw=3.3pt
    #1$}}%
    \fontdimen16\@aidxte=\@aidxtexvi
    \fontdimen17\@aidxte=\@aidxtexvii
    \fontdimen16\@aidxel=\@aidxelxvi
    \fontdimen17\@aidxel=\@aidxelxvii
    \fontdimen16\@aidxtw=\@aidxtwxvi
    \fontdimen17\@aidxtw=\@aidxtwxvii}
\makeatother

\def\eqref#1{(\ref{#1})}%

\def\gl{{\mathfrak{gl}}}
\def\ssl{{\mathfrak{sl}}}

\def\ie{{\sl i.e.\/}}
\def\eg{{\sl e.g.\/}}

\def\etc{{\sl etc.\/}}
\def\cf{{\sl cf.\/}}

\def\nn{\notag}

\def\C{{\mathbbm C}}
\def\N{{\mathbbm N}}

\version{17 November 2000}%
\preprint{DAMTP-2000-127}%

\def\q#1{{[#1]}_q}%
\def\Ha{H\otimes\openone}
\def\Hb{\openone\otimes H}
\def\Dop{\Delta^{\rm op}}
\def\dfrac#1#2{\frac{\displaystyle #1}{\displaystyle #2}}
\def\dd{\delta_1-\delta_2}%
\def\qfrac#1#2{\frac{\scriptstyle \q{#1}}{\scriptstyle \q{#2}}}%
\def\sfrac#1#2{\frac{\scriptstyle #1}{\scriptstyle #2}}%
\def\kk#1#2{\ket{#1}\otimes\ket{#2}}%

\begin{document}
%

\title{\dopreprint Factorizing twists and $R$-matrices for representations\\
                   of the quantum affine algebra $U_q(\hat\ssl_2)$}
\author{Hendryk Pfeiffer\thanks{e-mail: H.Pfeiffer@damtp.cam.ac.uk}}
\address{Department of Applied Mathematics and Theoretical Physics,\\
  Centre for Mathematical Sciences,\\
  Wilberforce Road, Cambridge CB3 0WA, UK}
\date{\version}
\maketitle

%
\begin{abstract}
%

  We calculate factorizing twists in evaluation representations of the
  quantum affine algebra $U_q(\hat\ssl_2)$. From the factorizing
  twists we derive a representation independent expression of the
  $R$-matrices of $U_q(\hat\ssl_2)$. Comparing with the corresponding
  quantities for the Yangian $Y(\ssl_2)$, it is shown that the
  $U_q(\hat\ssl_2)$ results can be obtained by `replacing numbers by
  $q$-numbers'. Conversely, the limit $q\rightarrow 1$ exists in
  representations of $U_q(\hat\ssl_2)$ and both the factorizing twists
  and the $R$-matrices of the Yangian $Y(\ssl_2)$ are recovered in
  this limit.

\end{abstract}


%
\section{Introduction}
%

The notion of twists and factorizing twists was introduced by
Drinfel'd~\cite{Dr90} in the context of quasi-Hopf algebras. They were
applied very successfully to the Hopf algebras used in the framework
of the Algebraic Bethe Ansatz for quantum integrable spin
chains. Maillet and Sanchez de Santos found factorizing twists in this
context for the fundamental evaluation representations of the Yangian
$Y(\ssl_2)$ and of the quantum affine algebra
$U_q(\hat\ssl_2)$~\cite{MaSa96}.

Such a factorizing twist $F$ equips the quantum algebra with a new
coproduct $\Delta_F=F\cdot\Delta\cdot F^{-1}$ which is
cocommutative. In representations the twist corresponds to a change of
basis on the tensor product which is in general not induced by changes
of basis of the tensor factors. In the new basis, the coproduct
then becomes invariant under permutations of the tensor factors and
is cocommutative. 

The cocommutativity of the coproduct has provided a dramatic
simplification of the Algebraic Bethe Ansatz. The application of
factorizing twists to this ansatz has triggered a rapid development
encompassing the solution of the inverse problem, the calculation of
correlation functions (see \eg~\cite{KiMa99,IzKi99,KiMa00}) and a
simplification of the nesting procedure if algebras of higher
rank are considered~\cite{AlBo00a}.

Factorizing twists were constructed first for the fundamental
evaluation representations of the Yangian $Y(\ssl_2)$ and of the
quantum affine algebra $U_q(\hat\ssl_2)$~\cite{MaSa96}. This was
recently extended to arbitrary evaluation representations of the
Yangian $Y(\ssl_2)$~\cite{Pf00a,Te99} and to the fundamental
representation of the Yangian $Y(\ssl_n)$ for general
$n$~\cite{AlBo00a}.

In the present paper, we derive factorizing twists in
finite-dimensional evaluation representations of the quantum affine
algebra $U_q(\hat\ssl_2)$ in a direct calculation ($q$ not a root of
unity). It is possible to follow the approach presented
in~\cite{Pf00a} and to generalize the results from the Yangian
$Y(\ssl_2)$ almost literally to the quantum affine algebra
$U_q(\hat\ssl_2)$, replacing numbers by $q$-numbers. Conversely, the
limit $q\rightarrow 1$ exists in representations of $U_q(\hat\ssl_2)$
both for the twist and for the $R$-matrix, and the corresponding
Yangian results are recovered. This limit $q\rightarrow 1$
demonstrates explicitly the close relation of the two algebras at the
representation level. One might have conjectured such a result from
the fact that the quantum affine algebra essentially specializes to
the Yangian if $q=1$; see, for example, the comments
in~\cite{Dr87,ChPr94}. However, this relation has not been seen
explicitly at the representation level before.

Finally, the $R$-matrix for generic evaluation representations is 
determined in its Gauss decomposition. In particular, no fusion is
necessary to determine the $R$-matrices of higher representations.

The key idea of our calculations is to choose a suitable presentation
of the evaluation representations of $U_q(\hat\ssl_2)$ which exhibits
close similarities with the Yangian $Y(\ssl_2)$, namely compared with
the Yangian, numbers are replaced by $q$-numbers. Then the
machinery developed in~\cite{Pf00a} can be applied in a
straightforward way. A remarkable result is that all intermediate
steps in the calculation of the twists for the Yangian hold for
$q$-numbers as well and thus generalize almost literally to
$U_q(\hat\ssl_2)$. This reveals the deep similarity of
$U_q(\hat\ssl_2)$ and $Y(\ssl_2)$ at the representation level.

At present the form of the factorizing twists on the abstract algebra
$U_q(\hat\ssl_2)$ is not known. This is partly due to the fact that
the twists have been first discovered in representations rather than
on the algebra, and it is not possible to just pull back the coproduct
along the evaluation homomorphism $U_q(\hat\ssl_2)\to U_q(\ssl_2)$
because this map does not preserve the co-algebra
structure. Furthermore, one does not expect a factorizing twist as a
proper element of $U_q(\hat\ssl_2)\otimes U_q(\hat\ssl_2)$, but only
as a formal expression which makes sense on a class of
representations which is determined in Section~\ref{sect_exist}.

The present paper is organized as follows. In Section~\ref{sect_prep}
we recall the basic properties of factorizing twists and introduce our
notation for the study of evaluation representations of the quantum
affine algebra $U_q(\hat\ssl_2)$. The factorizing twists in a generic
evaluation representation are then constructed in
Section~\ref{sect_twist}. The calculations follow closely the lines
of~\cite{Pf00a}, and we keep the presentation of the results very
brief. The precise conditions for the existence of the twists in
finite-dimensional evaluation representations are given in
Section~\ref{sect_exist}. In Section~\ref{sect_rmatrix} we construct
$R$-matrices in evaluation representations of
$U_q(\hat\ssl_2)$. Examples of factorizing twists are finally
tabulated in Section~\ref{sect_examples}.

%
\section{Preliminaries}
%
\label{sect_prep}

\subsection{Drinfel'd twists}

First we recall the definition and key properties of factorizing
twists. They are due to Drinfel'd~\cite{Dr87,Dr90} where more details
can be found. We write $\mu,\eta,\Delta,\epsilon,S$ for the product,
unit, coproduct, co-unit and antipode of a Hopf algebra.

\begin{definition}
A Hopf algebra $\sym{A}$ is called \emph{quasi-triangular} if there
exists an invertible element $R\in\sym{A}\otimes\sym{A}$, called the
\emph{universal $R$-matrix}, which satisfies
\begin{mathletters}
\begin{eqnarray}
\label{eq_nonlineara}
  (\Delta\otimes\id)(R) &=& R_{13}R_{23},\\
\label{eq_nonlinearb}
  (\id\otimes\Delta)(R) &=& R_{13}R_{12},\\
\label{eq_almostcoc}
  \Dop(a) &=& R\cdot\Delta(a)\cdot R^{-1},
\end{eqnarray}%
\end{mathletters}%
for all $a\in\sym{A}$. The Hopf algebra $\sym{A}$ is called
\emph{triangular} if in addition
\begin{equation}
  \alignidx{R_{21}=R_{12}^{-1}}.
\end{equation}
Here $R_{ij}$ denote as usual the embeddings of $R$ in the different
factors of $\sym{A}\otimes\sym{A}\otimes\sym{A}$.
\end{definition}

\begin{definition}
Let $\sym{A}$ be a Hopf algebra. An invertible element
$F\in\sym{A}\otimes\sym{A}$ is called a \emph{co-unital $2$-cocycle} or
\emph{Drinfel'd twist} if it satisfies
\begin{mathletters}
\begin{eqnarray}
  (\epsilon\otimes\id)(F) &=& 1,\\
  (\id\otimes\epsilon)(F) &=& 1,\\
\label{eq_cocycle}
  F_{12}\cdot (\Delta\otimes\id)(F) &=& F_{23}\cdot (\id\otimes\Delta)(F).
\end{eqnarray}%
\end{mathletters}%
\end{definition}

\begin{theorem}[Drinfel'd]
\label{thm_twist}
Let $\sym{A}$ be a quasi-triangular Hopf algebra and
$F\in\sym{A}\otimes\sym{A}$ be a co-unital $2$-cocycle. Then the
algebra of $\sym{A}$ together with the operations
\begin{mathletters}
\begin{eqnarray}
  \Delta_F(a) &:=& F\cdot\Delta(a)\cdot F^{-1},\\
\label{eq_antipode}
  S_F(a) &:=& u\cdot S(a)\cdot u^{-1},\qquad u:= \mu(\id\otimes S)(F),\\
  R_F &:=& F_{21}\cdot R\cdot F^{-1}
\end{eqnarray}%
\end{mathletters}%
and the old co-unit $\epsilon$, forms a quasi-triangular Hopf algebra
$\sym{A}_F$. 
\end{theorem}

The cocycle condition~\eqref{eq_cocycle} is required in order to make
the twisted coproduct $\Delta_F$ co-associative so that one obtains a
Hopf algebra $\sym{A}_F$ rather than just a quasi-Hopf algebra.

\begin{definition}
Let $\sym{A}$ be a quasi-triangular Hopf algebra with a co-unital
$2$-cocycle $F\in\sym{A}\otimes\sym{A}$. $F$ is called a
\emph{factorizing twist} if $R_F=1\otimes 1$ in
Theorem~\ref{thm_twist}, \ie\
\begin{equation}
  \alignidx{R_{12}=F_{21}^{-1}\cdot F_{12}}.
\end{equation}
\end{definition}

\begin{remark}
\label{rem_opposite}
\begin{myenumerate}
\item
  If a quasi-triangular Hopf algebra $\sym{A}$ admits a factorizing
  twist, the twisted coproduct $\Delta_F$ is cocommutative.
\item
  In this case the Hopf algebra is triangular since its universal
  $R$-matrix satisfies
\begin{equation}
  \alignidx{R_{21}=F_{12}^{-1}\cdot F_{21}=R_{12}^{-1}}.
\end{equation}
\item
  The converse implication is true at least for finite-dimensional
  semi-simple Hopf algebras~\cite{EtGe98a}. 
\item
  If $F$ is a factorizing twist, then the opposite coproduct can be
  made cocommutative using the twist $F_{21}$ rather than $F_{12}$:
\begin{equation}
  F_{21}\cdot\Dop(a)\cdot F_{21}^{-1} = \Delta_F(a)
    = F_{12}\cdot\Delta(a)\cdot F_{12}^{-1}.
\end{equation}
\end{myenumerate}
\end{remark}

In the discussion of the quantum affine algebra $U_q(\hat\ssl_2)$ we
face the difficulty that these algebras do not have a universal
$R$-matrix as an element of $\sym{A}\otimes\sym{A}$, but rather a
pseudo-universal $R$-matrix, \ie\ a formal expression which gives rise
to a corresponding quantity only on a certain class of representations
(in this case on the irreducible representations). It is furthermore
known~\cite{ChPr94} that the quantum affine algebra $U_q(\hat\ssl_2)$
is pseudo-triangular.

We thus expect only a pseudo-factorizing twist, \ie\ the twist should
exist for a large class of representations, but not be an element of
$\sym{A}\otimes\sym{A}$. In Section~\ref{sect_exist} we state precise
conditions for which evaluation representations our formal expressions
give rise to well-defined and invertible linear maps.

\subsection{Notations and conventions}

In this section we explain our notation for finite-dimensional
evaluation representations of the quantum affine algebra
$U_q(\hat\ssl_2)$. The central idea is to employ a notation for
$U_q(\hat\ssl_2)$ which shows sufficiently close similarities with the
Yangian $Y(\ssl_2)$ in order to apply the methods of~\cite{Pf00a}.

Similarly to the case of the Yangian $Y(\ssl_2)$ whose evaluation
representations can be described using representations of the Lie
algebra $\ssl_2$, it is possible to write the finite-dimensional
evaluation representations of the quantum affine algebra
$U_q(\hat\ssl_2)$ in terms of certain representations of the algebra
$U_q(\ssl_2)$ (without hat) \cite{ChPr91}. There are two versions of
these algebras, one in terms of formal power series in $q-1$, the
other with a special complex number $q$. In the latter case, the
representations have essentially the same structure only if $q$ is not
a root of unity.

The algebra $U_q(\ssl_2)$ is defined~\cite{Ka95a}, in terms of the
generators $E$, $F$, $K$, $K^{-1}$ and the relations,
\begin{mathletters}
\begin{gather}
  K\,K^{-1}=1=K^{-1}\,K,\\
  K\,E\,K^{-1} = q^2\,E,\\
  K\,F\,K^{-1} = q^{-2}\,F,\\
  [E,F] = \frac{K-K^{-1}}{q-q^{-1}},
\end{gather}%
\end{mathletters}%
for $q\notin\{-1,1\}$.

If $q$ is not a root of unity, its finite-dimensional irreducible
complex representations are given up to equivalence by the vector
spaces $V_{\epsilon,n}\cong\C^{n+1}$, $\epsilon\in\{-1,1\}$,
$n\in\N_0$. $U_q(\ssl_2)$ acts on a basis $\ket{0},\ldots,\ket{n}$ of
$V_{\epsilon,n}$ as follows,
\begin{mathletters}
\begin{eqnarray}
  E\ket{m} &=& \epsilon\q{m}\,\ket{m-1},\\
  F\ket{m} &=& \q{n-m}\,\ket{m+1},\\
  K\ket{m} &=& \epsilon\,q^{n-2m}\,\ket{m}.
\end{eqnarray}%
\end{mathletters}%
Here we have used the $q$-numbers
\begin{equation}
  \q{k} := \frac{q^k-q^{-k}}{q-q^{-1}}.
\end{equation}
In order to describe the evaluation representations of
$U_q(\hat\ssl_2)$, we need only the `type 1' representations $V_{1,n}$
for which $\epsilon=1$. There is a Casimir element
\begin{equation}
  C^{(2)} = \frac{qK+q^{-1}K^{-1}}{{(q-q^{-1})}^2} + FE,
\end{equation}          
whose eigenvalue on $V_{1,n}$ is given by
\begin{equation}
  \frac{q^{n+1}+q^{-n-1}}{{(q-q^{-1})}^2}.
\end{equation}
It is convenient to cast these representations in a form which looks
more similar to the Lie algebra $\ssl_2$ in Cartan-Weyl
form. Therefore we define for each representation $V_{1,n}$ an
operator $H$ by
\begin{equation}
  H\ket{m} := (\frac{n}{2}-m)\,\ket{m},
\end{equation}
\ie\ $K=q^{2H}$ so that $E$, $F$ and $H$ satisfy the following
identities on representations $V_{1,n}$,
\begin{mathletters}
\label{eq_uhsl2}
\begin{eqnarray}
  [H,E] &=& E,\\\relax
  [H,F] &=& -F,\\\relax
  [E,F] &=& \q{2H}.
\end{eqnarray}%
\end{mathletters}%
In this formulation, the limit $q\rightarrow 1$ makes sense, and the
generators $E$, $F$ tend towards their $\ssl_2$ counterparts while $H$
remains unchanged. The vector $\ket{0}$ then corresponds to the
highest weight vector with weight $\frac{n}{2}$ of $\ssl_2$. The point
is here to illustrate the limit $q\rightarrow 1$ for given
representations rather than analyzing the representations of the
algebra~\eqref{eq_uhsl2}.

The finite-dimensional evaluation representations $V_\lambda(w)$ of
$U_q(\hat\ssl_2)$ are characterized by a representation
$V_{1,2\lambda}$ of $U_q(\ssl_2)$, $2\lambda\in\N_0$, $q$ not a root
of unity, and by a parameter $w\in\C$~\cite{ChPr91}. Here we use the
physicists' presentation of $U_q(\hat\ssl_2)$ in terms of generators
$T^i_j(z)$ which satisfy the Quantum Yang-Baxter-Equation in the form
of the $RTT$ relations
\begin{equation}
\label{eq_rtt}
  R^{ij}_{k\ell}(v/z)T^k_p(v)T^\ell_q(z) = T^j_\ell(z)T^i_k(v)R^{k\ell}_{pq}(v/z),
\end{equation}
with the trigonometric $R$-matrix
\begin{equation}
\label{eq_rtrig}
  R(z) = \begin{pmatrix}
           1\\
           &\dfrac{z-z^{-1}}{zq-{(zq)}^{-1}}&\dfrac{q-q^{-1}}{zq-{(zq)}^{-1}}\\
           &\dfrac{q-q^{-1}}{zq-{(zq)}^{-1}}&\dfrac{z-z^{-1}}{zq-{(zq)}^{-1}}\\
           &&&1
         \end{pmatrix}.
\end{equation}
The functionals $T^i_j(z)$ in~\eqref{eq_rtt} represent the generators
of $U_q(\hat\ssl_2)$ in the evaluation representation $V_\lambda(w)$.
They read in terms of the operators~\eqref{eq_uhsl2},
\begin{mathletters}
\label{eq_generatorsmult}
\begin{eqnarray}
  A(z) = T^1_1(z) &=& \frac{\frac{z}{w}q^{H+\frac{1}{2}}-\frac{w}{z}q^{-H-\frac{1}{2}}}{q-q^{-1}},\\
  B(z) = T^1_2(z) &=& F\\
  C(z) = T^2_1(z) &=& E\\
  D(z) = T^2_2(z) &=& \frac{\frac{z}{w}q^{-H+\frac{1}{2}}-\frac{w}{z}q^{H-\frac{1}{2}}}{q-q^{-1}}.
\end{eqnarray}%
\end{mathletters}%
Note that there are several versions of~\eqref{eq_rtrig}
and~\eqref{eq_generatorsmult} in the literature which are related by
changes of the basis that can depend on $z$. For a comparison with the
Yangian, the above choice is most convenient.

Finally, we write $z=q^u$, $w=q^\delta$ and $q=e^h$ so that the
spectral parameter becomes additive, and the trigonometric nature of
the $R$-matrix is apparent. We find the following expressions:
\begin{mathletters}
\begin{eqnarray}
  A(u) = T^1_1(z) &=& \q{u-\delta+H+\frac{1}{2}},\\
  B(u) = T^1_2(z) &=& F\\
  C(u) = T^2_1(z) &=& E\\
  D(u) = T^2_2(z) &=& \q{u-\delta-H+\frac{1}{2}}.
\end{eqnarray}%
\end{mathletters}%
In the following, we use this notation in terms of $q$-numbers in
order to make the similarities with the Yangian obvious. With this
notation, however, we always mean the corresponding expressions in
terms of $z$, $w$ and $q$ as given in~\eqref{eq_generatorsmult}.

Observe that in~\eqref{eq_generatorsmult} we have chosen a particular
normalization. With this choice, the quantum determinant is given by
\begin{equation}
  \qdet T(z) =
  \frac{\frac{z^2q}{w^2}+\frac{w^2}{z^2q}}{{(q-q^{-1})}^2} - C^{(2)},
\end{equation}
where $C^{(2)}$ denotes the Casimir element of $U_q(\ssl_2)$.  On
evaluation representations $V_\lambda(w)$ the above equation reads
\begin{equation}
\label{eq_qdetaff}
  \qdet T(z) =
  \frac{\frac{z^2q}{w^2}+\frac{w^2}{z^2q}}{{(q-q^{-1})}^2} -
  \frac{q^{2\lambda+1}+q^{-2\lambda-1}}{{(q-q^{-1})}^2}
  = \q{u-\delta+\lambda+1}\,\q{u-\delta-\lambda},
\end{equation}
where $z=q^u$ and $w=q^\delta$. Therefore, strictly speaking, we treat
$V_\lambda(w)$ as a representation of $U_q(\hat\gl_2)$.

For compatibility with~\cite{Te99,Pf00a}, we use the following
coproduct (which might be more naturally called the opposite
coproduct), 
\begin{mathletters}
\begin{eqnarray}
  \Delta A(u)&=&A(u)\otimes A(u)+C(u)\otimes B(u),\\
  \Delta C(u)&=&A(u)\otimes C(u)+C(u)\otimes D(u),\\
  \Delta B(u)&=&B(u)\otimes A(u)+D(u)\otimes B(u),\\
  \Delta D(u)&=&B(u)\otimes C(u)+D(u)\otimes D(u).
\end{eqnarray}%
\end{mathletters}%
This coproduct acts on $V_{\lambda_1}(w_1)\otimes V_{\lambda_2}(w_2)$
as
\begin{mathletters}
\begin{alignat}{2}
  \Delta A(u)\ket{\ell,k} &= a_\ell^{(1)}(u)\,a_k^{(2)}(u)\ket{\ell,k} 
    &+& \q{\ell}\q{2\lambda_2-k}\ket{\ell-1,k+1},\\
  \Delta B(u)\ket{\ell,k} &= a_k^{(2)}(u)\,\q{2\lambda_1-\ell}\ket{\ell+1,k}
    &+& d_\ell^{(1)}(u)\,\q{2\lambda_2-k}\ket{\ell,k+1},\\
  \Delta C(u)\ket{\ell,k} &= a_\ell^{(1)}(u)\,\q{k}\ket{\ell,k-1}
    &+& d_k^{(2)}(u)\,\q{\ell}\ket{\ell-1,k},\\
  \Delta D(u)\ket{\ell,k} &= \q{2\lambda_1-\ell}\q{k}\ket{\ell+1,k-1}
    &+& d_\ell^{(1)}(u)\,d_k^{(2)}(u)\,\ket{\ell,k},
\end{alignat}%
\end{mathletters}%
where we write $\ket{\ell,k}:=\ket{\ell}\otimes\ket{k}$, $w_j=q^{\delta_j}$ and
\begin{mathletters}
\begin{eqnarray}
  a_k^{(j)}(u) &:=& \q{u-\delta_j+\lambda_j-k+\frac{1}{2}},\\
  d_k^{(j)}(u) &:=& \q{u-\delta_j-\lambda_j+k+\frac{1}{2}}.
\end{eqnarray}%
\end{mathletters}%

\subsection{Irreducibility of evaluation representations of $U_q(\hat\ssl_2)$}

The following statements concerning evaluation representations of the
quantum affine algebra go back to the work of V.~Tarasov~\cite{Ta85b}
and can be found, for example, in~\cite{ChPr91}.

\begin{theorem}
Each finite-dimensional irreducible type $(1,1)$ representation of the
quantum affine algebra $U_q(\hat\ssl_2)$ over $\C$ is isomorphic to a
tensor product of evaluation representations. Two such tensor products
describe isomorphic representations if and only if they are related by
a permutation of the tensor factors.
\end{theorem}

\begin{theorem}
\label{thm_irrepaff}
The tensor product of finite-dimensional evaluation representations
$V_{\lambda_1}(w_1)\otimes V_{\lambda_2}(w_2)$ of $U_q(\hat\ssl_2)$ is
reducible if and only if
\begin{equation}
  \frac{w_1^2}{w_2^2} = q^{\pm2(\lambda_1+\lambda_2-j+1)},
\end{equation}  
for an integer $j$ satisfying $1\leq
j\leq\min\{2\,\lambda_1,2\,\lambda_2\}$. In this case the
representation is neither completely reducible nor isomorphic to
$V_{\lambda_2}(w_2)\otimes V_{\lambda_1}(w_1)$.
\end{theorem}

Note that we have adapted the above result to our notation which uses
a basis of Cartan-Weyl type.

%
\section{Construction of factorizing twists}
%
\label{sect_twist}

The strategy how to construct the factorizing twists is the
following. In each evaluation representation, the twist $F_{12}^{-1}$
is given by a lower triangular matrix. It is calculated in a form
decomposed into a diagonal matrix times a lower triangular matrix
which has only entries `$1$' on its diagonal. The triangular part of
the twist is the change of basis operator on the tensor product
$V_{\lambda_1}(w_1)\otimes V_{\lambda_2}(w_2)$ which diagonalizes
$\Delta D(u)$. The prefactors of the eigenvectors of $\Delta D(u)$
form the diagonal part of the twist. They are fixed in a second step
by the requirement that the twisted coproduct has to be cocommutative.

\subsection{The triangular part}

It is convenient to make use of the following abbreviations, \cf\ the
corresponding definitions for the Yangian in equations~(3.10c)
and~(3.16c) of~\cite{Pf00a}:
\begin{mathletters}
\begin{eqnarray}
\label{eq_gq}
  g(x) &:=& \q{\delta_1-\delta_2+\lambda_1-\lambda_2+x},\\
  \tilde g(x) &:=& \q{\delta_2-\delta_1+\lambda_1-\lambda_2+x}.
\end{eqnarray}%
\end{mathletters}%
Furthermore we define the $q$-factorial
$\q{n}!:=\q{n}\,\q{n-1}\cdots\q{1}$; $\q{0}!:=1$ for $n\in\N$.

Using the same method as in Section~3 of~\cite{Pf00a}, we can
diagonalize $\Delta D(u)$ on each representation
$V_{\lambda_1}(w_1)\otimes V_{\lambda_2}(w_2)$. The remarkable result
is that the structure of the solutions does not change compared with
the Yangian $Y(\ssl_2)$. The only difference is the appearance of the
$q$-version of $g(x)$~\eqref{eq_gq}.

\begin{lemma}
The eigenvectors of $\Delta D(u)$ on $V_{\lambda_1}(w_1)\otimes
V_{\lambda_2}(w_2)$ are independent of $u$. They are given by
\begin{equation}
\label{eq_diagonalized}
  v_{\ell k} = q_{\ell k}\sum_{n=0}^{\min\{k,2\lambda_1-\ell\}}
    \frac{{(-1)}^n}{\q{n}!}\,\Bigl(\prod_{j=1}^n
    \frac{\q{2\lambda_1-\ell-j+1}\q{k-j+1}}{g(k-\ell-j)}\Bigr)\,
    \ket{\ell+n,k-n}
\end{equation}
and correspond to the eigenvalue $d_\ell^{(1)}(u)\cdot d_k^{(2)}(u)$.
Here the $q_{\ell k}$ are arbitrary factors. The inverse
transformation is given by
\begin{equation}
\label{eq_diagonalizedi}
  \ket{\ell,k} = \sum_{n=0}^{\min\{k,2\lambda_1-\ell\}}
    \frac{1}{\q{n}!}\,\Bigl(\prod_{j=1}^n
    \frac{\q{2\lambda_1-\ell-j+1}\q{k-j+1}}{g((k-n)-(\ell+n)+j)}\Bigr)\,
    \frac{v_{\ell+n,k-n}}{q_{\ell+n,k-n}}.
\end{equation}
\end{lemma}

This change of basis determines the triangular part of the factorizing twist:
\begin{proposition}
The expression
\begin{equation}
\label{eq_f12i}
  F_{12}^{-1} = \biggl(\sum_{n=0}^\infty\frac{{(-1)}^n}{\q{n}!}
    F^n\otimes E^n\prod_{j=1}^n
    {\q{\delta_1-\delta_2+\Ha-\Hb-j}}^{-1}\biggr)\,Q_{12}^{-1},
\end{equation}
specializes to the change of basis~\eqref{eq_diagonalized} on all
representations $V_{\lambda_1}(w_1)\otimes V_{\lambda_2}(w_2)$. The
expression
\begin{equation}
\label{eq_f12}
  F_{12} = Q_{12}\sum_{n=0}^\infty\frac{1}{\q{n}!}
    \biggl(\prod_{j=1}^n
    {\q{\delta_1-\delta_2+\Ha-\Hb+j}}^{-1}\biggr)F^n\otimes E^n
\end{equation}
specializes to the inverse change of basis~\eqref{eq_diagonalizedi}. Here
\begin{equation}
  Q_{12}^{-1}\ket{\ell,k} = q_{\ell k}\ket{\ell,k}
\end{equation}
denotes the diagonal part of the twist.
\end{proposition}

In addition to this twist which diagonalizes $\Delta D(u)$ on
evaluation representations, it is possible to construct another twist
from the requirement that it diagonalizes $\Delta A(u)$.
\begin{lemma}
The eigenvectors of $\Delta A(u)$ on $V_{\lambda_1}(w_1)\otimes
V_{\lambda_2}(w_2)$ are independent of $u$. They are given by
\begin{equation}
\label{eq_diagonalizea}
  \tilde v_{\ell k} = \tilde q_{\ell k}\sum_{n=0}^{\min\{2\lambda_2-k,\ell\}}
    \frac{1}{\q{n}!}\,\Bigl(\prod_{j=1}^n
    \frac{\q{2\lambda_2-k-j+1}\q{\ell-j+1}}{\tilde g(k-\ell+j)}\Bigr)\,
    \ket{\ell-n,k+n}
\end{equation}
and correspond to the eigenvalues $a_\ell^{(1)}(u)\cdot
a_k^{(2)}(u)$. Here the $\tilde q_{\ell k}$ are arbitrary factors. The
inverse transformation is given by
\begin{equation}
\label{eq_diagonalizeai}
  \ket{\ell,k} = \sum_{n=0}^{\min\{2\lambda_2-k,\ell\}}
    \frac{{(-1)}^n}{\q{n}!}\,\Bigl(\prod_{j=1}^n
    \frac{\q{2\lambda_2-k-j+1}\q{\ell-j+1}}{\tilde g((k+n)-(\ell-n)-j)}\Bigr)\,
    \frac{\tilde v_{\ell-n,k+n}}{\tilde q_{\ell-n,k+n}}.
\end{equation}
\end{lemma}

This change of basis determines the triangular part of another
factorizing twist:
\begin{proposition}
The expressions
\begin{mathletters}
\begin{eqnarray}
\label{eq_f12ti}
  \tilde F_{12}^{-1}
    &=& \biggl(\sum_{n=0}^\infty\frac{1}{\q{n}!}E^n\otimes F^n\prod_{j=1}^n
                {\q{\delta_2-\delta_1+\Ha-\Hb+j}}^{-1}\biggr)\,\tilde Q_{12}^{-1},\\
  \tilde F_{12} 
    &=& \tilde Q_{12}\sum_{n=0}^\infty\frac{{(-1)}^n}{\q{n}!}\biggl(\prod_{j=1}^n
           {\q{\delta_2-\delta_1+\Ha-\Hb-j}}^{-1}\biggr)E^n\otimes F^n,
\end{eqnarray}
\end{mathletters}%
specialize to the change of basis~\eqref{eq_diagonalizea} and its
inverse transformation~\eqref{eq_diagonalizeai} on all representations
$V_{\lambda_1}(w_1)\otimes V_{\lambda_2}(w_2)$. 
Here 
\begin{equation}
  \tilde Q_{12}^{-1}\ket{\ell,k} = \tilde q_{\ell k}\ket{\ell,k}.
\end{equation}
denotes the diagonal part of the twist.
\end{proposition}

\subsection{The diagonal part}

At this point we know two different operators $F_{12}^{-1}$ and
$\tilde F_{12}^{-1}$ which provide a change of basis on the tensor
products of finite-dimensional evaluation representations
$V_{\lambda_1}(w_1)\otimes V_{\lambda_2}(w_2)$. The first one
diagonalizes $\Delta D(u)$ and the second $\Delta A(u)$. Their
diagonal parts $Q_{12}^{-1}$ and $\tilde Q_{12}^{-1}$ are yet
unspecified.

Employing the same method as in Section~3 of~\cite{Pf00a}, it can be
shown that a suitable choice of the diagonal part $Q_{12}^{-1}$ leads
to a cocommutative coproduct $\Delta_F = F_{12}\cdot\Delta\cdot
F_{12}^{-1}$. The required conditions on the diagonal part are stated
in the following propositions.

\begin{proposition}
\label{prop_newcopro}
If the coefficients $q_{\ell k}$ of $Q_{12}^{-1}$ satisfy the
recursion relations
\begin{equation}
\label{eq_recursd1}
  \frac{q_{\ell+1,k}}{q_{\ell k}} = \frac{g(-\ell-1)}{g(k-\ell-1)},\qquad
  \frac{q_{\ell,k+1}}{q_{\ell k}} = \frac{g(k-\ell)}{g(-2\lambda_1+k)},
\end{equation}
for $0\leq\ell\leq2\lambda_1$ and $0\leq k\leq2\lambda_2$, then the
twisted coproduct on $V_{\lambda_1}(w_1)\otimes V_{\lambda_2}(w_2)$ is
cocommutative. It has the form 
\begin{mathletters}
\begin{eqnarray}
  F_{12}\cdot\Delta D(u)\cdot F_{12}^{-1}
    &=& D(u)\otimes D(u),\\
  F_{12}\cdot\Delta B(u)\cdot F_{12}^{-1}
    &=& B(u)\otimes D(u)\,
                  \frac{\q{\delta_1-\delta_2+\Ha+\lambda_2}}
                       {\q{\delta_1-\delta_2+\Ha-\Hb}}\nn\\
    &&+ D(u)\otimes B(u)\,
                  \frac{\q{\delta_1-\delta_2-\lambda_1-\Hb}}
                       {\q{\delta_1-\delta_2+\Ha-\Hb}},\\
  F_{12}\cdot\Delta C(u)\cdot F_{12}^{-1}
     &=& C(u)\otimes D(u)\,
                  \frac{\q{\delta_1-\delta_2+\Ha-\lambda_2}}
                       {\q{\delta_1-\delta_2+\Ha-\Hb}}\nn\\
     &&+ D(u)\otimes C(u)\,
                  \frac{\q{\delta_1-\delta_2+\lambda_1-\Hb}}
                       {\q{\delta_1-\delta_2+\Ha-\Hb}}.
\end{eqnarray}%
\end{mathletters}%
\end{proposition}

\begin{proof}
The proof is almost literally the same as for Proposition~4.1
of~\cite{Pf00a}. The assertions reduce to the following identities for
$q$-numbers for $0\leq\ell\leq 2\lambda_1$, $0\leq k\leq
2\lambda_1$, $0\leq i\leq k$,
\begin{mathletters}
\label{eq_qidentities1}
\begin{gather}
  a_{k-i}^{(2)}(u)
    -d_{\ell+i+1}^{(1)}(u)\,
     \frac{\q{k-i}\q{2\lambda_2-k+i+1}}
          {\q{i+1}\,g(k-\ell-i-1)}\nn\\
  =d_k^{(2)}(u)\,
     \frac{g(2\lambda_2-\ell)\,g(-\ell-1)}
          {g(k-\ell)\,g(k-\ell-i-1)}
    -d_\ell^{(1)}(u)\,\frac{\q{k+1}\q{2\lambda_2-k}}{\q{i+1}\,g(k-\ell)},
\end{gather}
and for $1\leq i\leq k$,
\begin{gather}
  d_{k-i}^{(2)}(u)\,\q{\ell+i}\q{2\lambda_1-\ell-i+1}
    -a_{\ell+i-1}^{(1)}(u)\,\q{i}\,g(k-\ell-i)\nn\\
  =d_k^{(2)}(u)\,\q{\ell}\q{2\lambda_1-\ell+1}\,
      \frac{g(k-\ell-i)}{g(k-\ell)}
    -d_\ell^{(1)}(u)\,\q{i}\,
      \frac{g(k)\,g(-2\lambda_1+k-1)}{g(k-\ell)}.
\end{gather}%
\end{mathletters}%
It is a remarkable and non-trivial fact that these identities which
are known to hold without the $\q{\cdot}$-brackets, are still valid in this
form which involves the $q$-numbers.
\end{proof}

\begin{remark}
We can use the quantum determinant~\eqref{eq_qdetaff} in order to
prove that $F_{12}\cdot\Delta A(u)\cdot F_{12}^{-1}$ is also
cocommutative. 
\end{remark}

\begin{lemma}
The coefficients $q_{\ell k}$, given by
\begin{equation}
\label{eq_prodq}
  q_{\ell k} = \prod_{j=0}^{k-1}
     \frac{\q{\delta_1-\delta_2+\lambda_1-\lambda_2-\ell+j}}
          {\q{\delta_1-\delta_2-\lambda_1-\lambda_2+j}},
\end{equation}
satisfy the recursion relations~\eqref{eq_recursd1}.
\end{lemma}

In Section~5.3 of~\cite{Pf00a}, a second choice of the diagonal part
gave rise to a coproduct with a different ordering of operators. This
construction has also a correspondence in $U_q(\hat\ssl_2)$:

\begin{proposition}
\label{prop_newcopro2}
Let $\hat F_{12}^{-1}$ denote the expression~\eqref{eq_f12i} with a
diagonal part $\hat Q_{12}^{-1}$ rather than $Q_{12}^{-1}$. If the
coefficients $\hat q_{\ell k}$ of $\hat Q_{12}^{-1}$ satisfy
\begin{equation}
\label{eq_recursd2}
  \frac{\hat q_{\ell+1,k}}{\hat q_{\ell k}}
    = \frac{g(2\lambda_2-\ell)}{g(k-\ell)},\qquad
  \frac{\hat q_{\ell,k+1}}{\hat q_{\ell k}}
    = \frac{g(k-\ell+1)}{g(k+1)},
\end{equation}
for $0\leq\ell\leq2\lambda_1$, $0\leq k\leq2\lambda_2$, then the
twisted coproduct is also cocommutative. It is then given by
\begin{mathletters}
\begin{eqnarray}
  \hat F_{12}\cdot\Delta D(u)\cdot{\hat F_{12}}^{-1} 
    &=& D(u)\otimes D(u),\\
  \hat F_{12}\cdot\Delta B(u)\cdot{\hat F_{12}}^{-1} 
    &=& \frac{\q{\delta_1-\delta_2+\Ha-\lambda_2}}
             {\q{\delta_1-\delta_2+\Ha-\Hb}}\,
                  B(u)\otimes D(u)\nn\\
    &&+ \frac{\q{\delta_1-\delta_2+\lambda_1-\Hb}}
             {\q{\delta_1-\delta_2+\Ha-\Hb}}\,
                  D(u)\otimes B(u),\\
  \hat F_{12}\cdot\Delta C(u)\cdot{\hat F_{12}}^{-1} 
    &=& \frac{\q{\delta_1-\delta_2+\Ha+\lambda_2}}
             {\q{\delta_1-\delta_2+\Ha-\Hb}}\,
                  C(u)\otimes D(u)\nn\\
    &&+ \frac{\q{\delta_1-\delta_2-\lambda_1-\Hb}}
             {\q{\delta_1-\delta_2+\Ha-\Hb}}\,
                  D(u)\otimes C(u).
\end{eqnarray}%
\end{mathletters}%
\end{proposition}

\noindent
The proof of this proposition reduces to the same
identities~\eqref{eq_qidentities1}. 

\begin{lemma}
The coefficients $\hat q_{\ell k}$, given by
\begin{equation}
\label{eq_prodhq}
  \hat q_{\ell k} = \prod_{j=1}^\ell
     \frac{\q{\delta_1-\delta_2+\lambda_1+\lambda_2-\ell+j}}
          {\q{\delta_1-\delta_2+\lambda_1-\lambda_2+k-\ell+j}},
\end{equation}
satisfy the recursion relations~\eqref{eq_recursd2}.
\end{lemma}
Finally, analogous constructions are available for the twist
$\tilde F_{12}^{-1}$ which diagonalizes $\Delta A(u)$ rather than
$\Delta D(u)$:

\begin{proposition}
If the coefficients $\tilde q_{\ell k}$ of the diagonal part $\tilde
Q_{12}^{-1}$ of the transformation $\tilde F_{12}^{-1}$
in~\eqref{eq_f12ti} satisfy
\begin{equation}
\label{eq_recursa1}
  \frac{\tilde q_{\ell+1,k}}{\tilde q_{\ell k}}
  = \frac{\tilde g(k-\ell)}{\tilde g(2\lambda_2-\ell)},\qquad
  \frac{\tilde q_{\ell k}}{\tilde q_{\ell,k-1}}
  = \frac{\tilde g(k)}{\tilde g(k-\ell)},
\end{equation}
for $0\leq\ell\leq2\lambda_1$ and $0\leq k\leq2\lambda_2$, then the
twisted coproduct is cocommutative and given by
\begin{mathletters}
\begin{eqnarray}
  \tilde F_{12}\cdot\Delta A(u)\cdot{\tilde F_{12}}^{-1} 
    &=& A(u)\otimes A(u),\\
  \tilde F_{12}\cdot\Delta B(u)\cdot{\tilde F_{12}}^{-1} 
    &=& B(u)\otimes A(u)\,
                  \frac{\q{\delta_2-\delta_1+\Ha+\lambda_2}}
                       {\q{\delta_2-\delta_1+\Ha-\Hb}}\nn\\
              &&+ A(u)\otimes B(u)\,
                  \frac{\q{\delta_2-\delta_1-\lambda_1-\Hb}}
                       {\q{\delta_2-\delta_1+\Ha-\Hb}},\\
  \tilde F_{12}\cdot\Delta C(u)\cdot{\tilde F_{12}}^{-1} 
    &=& C(u)\otimes A(u)\,
                  \frac{\q{\delta_2-\delta_1+\Ha-\lambda_2}}
                       {\q{\delta_2-\delta_1+\Ha-\Hb}}\nn\\
              &&+ A(u)\otimes C(u)\,
                  \frac{\q{\delta_2-\delta_1+\lambda_1-\Hb}}
                       {\q{\delta_2-\delta_1+\Ha-\Hb}}.
\end{eqnarray}%
\end{mathletters}%
\end{proposition}

\begin{proof}
The proof of this proposition is similar to that of
Proposition~\ref{prop_newcopro}. It reduces to the following true
identities for $q$-numbers where $0\leq\ell\leq2\lambda_1$, $0\leq
k\leq2\lambda_2$ and $0\leq i\leq\ell$,
\begin{mathletters}
\label{eq_qidentities2}
\begin{gather}
  a_{k+i}^{(2)}(u)\,\q{\ell-i+1}\q{2\lambda_1-\ell+i}
    + d_{\ell-i+1}^{(1)}(u)\,\q{i}\,\tilde g(k-\ell+i)\nn\\
  = a_k^{(2)}(u)\,\q{\ell+1}\q{2\lambda_1-\ell}\,
        \frac{\tilde g(k-\ell+i)}{\tilde g(k-\ell)}
    + a_\ell^{(1)}(u)\,\q{i}\,
        \frac{\tilde g(-2\lambda_1+k)\,\tilde g(k+1)}{\tilde
    g(k-\ell)},
\end{gather}
and for $1\leq i\leq\ell$,
\begin{gather}
  a_{\ell-i}^{(1)}(u)\,\q{k+i}\q{2\lambda_2-k-i+1}
    + d_{k+i-1}^{(2)}(u)\,\q{i}\tilde g(k-\ell+i)\nn\\
  = a_k^{(2)}(u)\,\q{i}\,\frac{\tilde g(-\ell)\,\tilde g(2\lambda_2-\ell+1)}{\tilde g(k-\ell)}
    + a_\ell^{(1)}(u)\,\q{k}\q{2\lambda_2-k+1}\,\frac{\tilde g(k-\ell+1)}{\tilde g(k-\ell)}.
\end{gather}%
\end{mathletters}%
\end{proof}

\begin{lemma}
The coefficients $\tilde q_{\ell k}$,
\begin{equation}
\label{eq_prodtq}
  \tilde q_{\ell k}=\prod_{j=1}^\ell
    \frac{\q{\delta_2-\delta_1+\lambda_1-\lambda_2+k-\ell+j}}
         {\q{\delta_2-\delta_1+\lambda_1+\lambda_2-\ell+j}},
\end{equation}
satisfy the recursion formulas~\eqref{eq_recursa1}.
\end{lemma}

Again, there is an alternative choice for the diagonal part of the twist:
\begin{proposition}
Let ${\hat{\tilde F}}_{12}^{-1}$ denote same operator as $\tilde
F_{12}^{-1}$ in~\eqref{eq_f12ti}, but with a diagonal part given by
${\hat{\tilde Q}}_{12}^{-1}$. If the coefficients ${\hat{\tilde
q}}_{\ell k}$ of ${\hat{\tilde Q}}_{12}^{-1}$ satisfy
\begin{equation}
\label{eq_recursa2}
  \frac{\hat{\tilde q}_{\ell+1,k}}{\hat{\tilde q}_{\ell k}}
    = \frac{\tilde g(k-\ell-1)}{\tilde g(-\ell-1)},\qquad
  \frac{\hat{\tilde q}_{\ell k}}{\hat{\tilde q}_{\ell,k-1}}
    = \frac{\tilde g(-2\lambda_1+k-1)}{\tilde g(k-\ell-1)},
\end{equation}
for $0\leq\ell\leq2\lambda_1$, $0\leq k\leq2\lambda_2$, then the
twisted coproduct is cocommutative. It is of the form
\begin{mathletters}
\begin{eqnarray}
  \hat{\tilde F}_{12}\cdot\Delta A(u)\cdot{\hat{\tilde F}_{12}}^{-1} 
    &=& A(u)\otimes A(u),\\
  \hat{\tilde F}_{12}\cdot\Delta B(u)\cdot{\hat{\tilde F}_{12}}^{-1} 
    &=& \frac{\q{\delta_2-\delta_1+\Ha-\lambda_2}}
             {\q{\delta_2-\delta_1+\Ha-\Hb}}\,
                  B(u)\otimes A(u)\nn\\
    &&+ \frac{\q{\delta_2-\delta_1+\lambda_1-\Hb}}
             {\q{\delta_2-\delta_1+\Ha-\Hb}}\,
                  A(u)\otimes B(u),\\
  \hat{\tilde F}_{12}\cdot\Delta C(u)\cdot{\hat{\tilde F}_{12}}^{-1} 
    &=& \frac{\q{\delta_2-\delta_1+\Ha+\lambda_2}}
             {\q{\delta_2-\delta_1+\Ha-\Hb}}\,
                  C(u)\otimes A(u)\nn\\
    &&+ \frac{\q{\delta_2-\delta_1-\lambda_1-\Hb}}
             {\q{\delta_2-\delta_1+\Ha-\Hb}}\, 
                  A(u)\otimes C(u).
\end{eqnarray}%
\end{mathletters}%
\end{proposition}

\noindent
The proof is again similar and reduces to the same
identities~\eqref{eq_qidentities2}. 

\begin{lemma}
The coefficients ${\hat{\tilde q}}_{\ell k}$, given by
\begin{equation}
\label{eq_prodhtq}
  \hat{\tilde q}_{\ell k}=\prod_{j=0}^{k-1}
    \frac{\q{\delta_2-\delta_1-\lambda_1-\lambda_2+j}}
         {\q{\delta_2-\delta_1+\lambda_1-\lambda_2-\ell+j}},
\end{equation}
satisfy the recursion relations~\eqref{eq_recursa2}.
\end{lemma}

For the Yangian it was possible~\cite{Pf00a} to express the
diagonal parts of the various twists as quotients of Gamma
functions. The finite products like~\eqref{eq_prodq},
\eqref{eq_prodhq}, \eqref{eq_prodtq} and~\eqref{eq_prodhtq} then
appear in the specialization on evaluation representations when the
arguments of the Gamma functions of numerator and denominator differ
precisely by an integer number.

Here a similar construction is available which makes use of a
$q$-deformed Gamma function. There are several versions of such
functions defined in the literature. We need here the `modified
$q$-Gamma function' $\Gamma_q(t)$ (which is called $\hat\Gamma_q$
in~\cite{KlSc97}). It is characterized by the property
\begin{equation}
  \frac{\Gamma_q(t+1)}{\Gamma_q(t)} = \q{t} = \frac{q^t-q^{-t}}{q-q^{-1}}.
\end{equation}

\begin{proposition}
On each weight vector $\ket{\ell,k}$ of the representation
$V_{\lambda_1}(w_1)\otimes V_{\lambda_2}(w_2)$, the expressions
\begin{mathletters}
\label{eq_q12i}
\begin{eqnarray}
  Q_{12}^{-1} &=&
    \frac{\Gamma_q\bigl(\delta_1-\delta_2+\Ha-\Hb\bigr)}
         {\Gamma_q\bigl(\delta_1-\delta_2+\Ha-\lambda_2\bigr)}\,
    \frac{\Gamma_q\bigl(\delta_1-\delta_2-\lambda_1-\lambda_2\bigr)}
         {\Gamma_q\bigl(\delta_1-\delta_2-\lambda_1-\Hb\bigr)},\\
  \hat Q_{12}^{-1} &=&
    \frac{\Gamma_q\bigl(\delta_1-\delta_2+\Ha-\Hb+1\bigr)}
         {\Gamma_q\bigl(\delta_1-\delta_2+\Ha+\lambda_2+1\bigr)}\,
    \frac{\Gamma_q\bigl(\delta_1-\delta_2+\lambda_1+\lambda_2+1\bigr)}
         {\Gamma_q\bigl(\delta_1-\delta_2+\lambda_1-\Hb+1\bigr)},
\end{eqnarray}%
\end{mathletters}%
have the eigenvalues $q_{\ell k}$ and $\hat q_{\ell k}$ given
in~\eqref{eq_prodq} and~\eqref{eq_prodhq}. Furthermore we find
\begin{mathletters}
\label{eq_variousq}
\begin{eqnarray}
  \hat Q_{21}      &=& Q_{12}^{-1},\\
  \hat Q_{12}^{-1} &=& Q_{21},\\
  \tilde Q_{12}^{-1}         &=& \left.Q_{21}^{-1}\right|_{\delta_1\leftrightarrow\delta_2},\\
  {\hat{\tilde Q}_{12}}^{-1} &=& \left.Q_{12}\right|_{\delta_1\leftrightarrow\delta_2}.
\end{eqnarray}%
\end{mathletters}%
\end{proposition}

\begin{remark}
Observe that whenever the expressions for the factorizing twists
$F_{12}^{-1}$, $Q_{12}^{-1}$ \etc\ are written in a particular
finite-dimensional evaluation representation, the limit $q\rightarrow
1$ is well-defined. In this limit, the operators $E$ and $F$ of
$U_q(\ssl_2)$ tend towards their $\ssl_2$ counterparts.  On weight
vectors, the quotients of $q$-Gamma functions reduce to finite
products like~\eqref{eq_prodq}. Each factor of these products has a
well defined $q\rightarrow 1$ limit in which $q$-numbers tend towards
ordinary numbers. 

In all cases, the corresponding results for the Yangian $Y(\ssl_2)$
are recovered as a comparison with~\cite{Pf00a} reveals. The Yangian
results appear in the form where $\eta=1$ in the notation
of~\cite{Pf00a}.
\end{remark}

Example matrices $F_{12}^{-1}$ in finite-dimensional evaluation
representations of low dimension are tabulated in
Section~\ref{sect_examples}.

\subsection{Existence of the twist in representations}
\label{sect_exist}

Finally, we study for which finite-dimensional evaluation
representations $V_{\lambda_1}(w_1)\otimes V_{\lambda_2}(w_2)$ the
twist and its inverse exist. We restrict ourselves to the twist
$F_{12}^{-1}$ and its inverse $F_{12}$, \cf~\eqref{eq_f12i}
and~\eqref{eq_f12}.

Therefore, we have to write the coefficients of $Q_{12}^{-1}$ and
$Q_{12}$ in a form in which numerator and denominator have no common
factor. Furthermore we have to take into account that the numerator of
the diagonal part might cancel factors from the denominator of the
triangular part. The structure of the expressions is the same as for
the Yangian $Y(\ssl_2)$ in Section~5.4 of~\cite{Pf00a}:

\begin{proposition}
The expression $F_{12}^{-1}$, see equation~\eqref{eq_f12i}, is
well-defined in the representation $V_{\lambda_1}(w_1)\otimes
V_{\lambda_2}(w_2)$ if and only if
\begin{equation}
  \frac{w_1^2}{w_2^2}\neq q^{2(\lambda_1+\lambda_2-j+1)}
\end{equation}
for all integers $j$ in the range $1\leq j\leq\min\{2\lambda_1,2\lambda_2\}$.
\end{proposition}

\begin{proof}
From an analogous calculation as for the Yangian (Section~5.4
of~\cite{Pf00a}) we obtain a well-defined $F_{12}^{-1}$ if and only if
\begin{equation}
  \q{\delta_1-\delta_2-\lambda_1-\lambda_2+j-1}\neq 0
\end{equation}
for all integers $j$ in the range $1\leq
j\leq\min\{2\lambda_1,2\lambda_2\}$. We recall that
$w_j=q^{\delta_j}$, and the assertion follows.
\end{proof}

\begin{remark}
According to Theorem~\ref{thm_irrepaff}, $F_{12}^{-1}$ thus exists for
all finite-dimensional irreducible evaluation representations of
$U_q(\hat\ssl_2)$. 
\end{remark}

\begin{proposition}
The expression $F_{12}$ in~\eqref{eq_f12} is well-defined in the
representation $V_{\lambda_1}(w_1)\otimes V_{\lambda_2}(w_2)$ if and
only if
\begin{equation}
  \frac{w_1^2}{w_2^2}\neq q^{2(\lambda_1+\lambda_2-j+1)}
\end{equation}
for all integers $j$ in the range $2\leq j\leq2\lambda_1+2\lambda_2$.
\end{proposition}

\begin{remark}
There exist irreducible representations $V_{\lambda_1}(w_1)\otimes
V_{\lambda_2}(w_2)$ of $U_q(\hat\ssl_2)$ for which $F_{12}$ is not
well-defined. The situation is essentially the same as for the Yangian
$Y(\ssl_2)$.
\end{remark}

%
\section{$R$-matrices}
%
\label{sect_rmatrix}

From the factorizing twist which was determined in a representation
independent fashion, it is possible to derive a representation
independent $R$-matrix. This $R$-matrix appears automatically in a
canonical form, being Gauss decomposed into an upper triangular times
a diagonal times a lower triangular part.

We obtain the following expression for the $R$-matrix on
$V_{\lambda_1}(w_1)\otimes V_{\lambda_2}(w_2)$,
\begin{equation}
  R_{12} = \alignidx{F_{21}^{-1}\cdot F_{12} = R_+\cdot R_0\cdot R_-}.
\end{equation}
Here the upper triangular part $R_+$ is the triangular part of
$F_{21}^{-1}$ and can be obtained from~\eqref{eq_f12i}. The diagonal
part is the product $R_0=\alignidx{Q_{21}^{-1}\cdot Q_{12}}$,
\cf~\eqref{eq_q12i}, and the lower triangular part $R_-$ can be read
off from~\eqref{eq_f12},
\begin{mathletters}
\begin{eqnarray}
  R_+ &=& \sum_{n=0}^\infty\frac{1}{\q{n}!}
          E^n\otimes F^n\prod_{j=1}^n
          {\q{\delta_1-\delta_2+\Ha-\Hb+j}}^{-1},\\
  R_0 &=& \frac{\Gamma_q\bigl(\delta_1-\delta_2+\Ha+\lambda_2+1\bigr)}
               {\Gamma_q\bigl(\delta_1-\delta_2+\Ha-\Hb+1\bigr)}\cdot
          \frac{\Gamma_q\bigl(\delta_1-\delta_2+\lambda_1-\Hb+1\bigr)}
               {\Gamma_q\bigl(\delta_1-\delta_2+\lambda_1+\lambda_2+1\bigr)}\nn\\
      &&\times
          \frac{\Gamma_q\bigl(\delta_1-\delta_2+\Ha-\lambda_2\bigr)}
               {\Gamma_q\bigl(\delta_1-\delta_2+\Ha-\Hb\bigr)}\cdot
          \frac{\Gamma_q\bigl(\delta_1-\delta_2-\lambda_1-\Hb\bigr)}
               {\Gamma_q\bigl(\delta_1-\delta_2-\lambda_1-\lambda_2\bigr)},\\
  R_- &=& \sum_{n=0}^\infty\frac{1}{\q{n}!}
          \biggl(\prod_{j=1}^n {\q{\delta_1-\delta_2+\Ha-\Hb+j}}^{-1}\biggr)F^n\otimes E^n.
\end{eqnarray}%
\end{mathletters}%
The alternative twist $\hat F_{12}$ in
Proposition~\ref{prop_newcopro2} factorizes the same $R$-matrix,
\begin{equation}
  \alignidx{\hat F_{21}^{-1}\cdot\hat F_{12}=R_{12}},
\end{equation}
because the two factors of the diagonal part are just exchanged,
\cf~\eqref{eq_variousq}. The twist $\tilde F_{12}^{-1}$ which diagonalizes
$\Delta A(u)$, \cf~\eqref{eq_f12ti}, is related to $F_{12}^{-1}$ by
\begin{equation}
  \tilde F_{12}^{-1} = {\left.F_{21}^{-1}\right|}_{\delta_1\leftrightarrow\delta_2}.
\end{equation}
It thus factorizes the opposite $R$-matrix, but with negative spectral
parameter,
\begin{equation}
  \alignidx{\tilde F_{21}^{-1}\cdot\tilde F_{12}} =
  {\left.R_{21}\right|}_{\delta_1\leftrightarrow\delta_2}. 
\end{equation}

From the calculation of the $R$-matrix via the factorizing twists,
only the property of almost-cocommutativity~\eqref{eq_almostcoc} is
guaranteed by construction. However, this requirement has already
determined the $R$-matrix up to a constant representation dependent
factor. Recall that our $R$ is normalized such that
$R\ket{0}\otimes\ket{0}=\ket{0}\otimes\ket{0}$. Therefore it differs
from an $R$ matrix derived via the (pseudo-)universal $R$-matrix by a
multiplicative factor which is the character of the $R$-matrix. This
situation is again completely analogous to the Yangian.

%
\section{Examples}
%
\label{sect_examples}

In this section, we tabulate factorizing twists for some
finite-dimensional evaluation representations of low dimension.  We
list the matrices of the twists $F_{12}^{-1}$ in a weight basis. The
other twists and the $R$-matrices can then be calculated from
$F_{12}^{-1}$. The ordering of the basis vectors of
$V_{\lambda_1}(\delta_1)\otimes V_{\lambda_2}(\delta_2)$ is
$\ket{0}\otimes\ket{0},\ket{0}\otimes\ket{1},\ldots$.

Recall that due to the similarity of $U_q(\hat\ssl_2)$ with the
Yangian $Y(\ssl_2)$, one obtains twists for the Yangian if all
$q$-brackets in the following formulas are ignored.

\noindent
$V_\frac{1}{2}(\delta_1)\otimes V_\frac{1}{2}(\delta_2)$:
\begin{equation}
  F_{12}^{-1}=
  \begin{pmatrix}
    1\\
    & \qfrac{\dd}{\dd-1} \\
    & -\frac{1}{\q{\dd-1}} & 1 \\
    &&& 1
  \end{pmatrix}
\end{equation}

\noindent
$V_\frac{1}{2}(\delta_1)\otimes V_1(\delta_2)$:
\begin{equation}
  F_{12}^{-1}=
  \begin{pmatrix}
    1\\
    &\qfrac{\dd-\frac{1}{2}}{\dd-\frac{3}{2}}\\
    &0&\qfrac{\dd+\frac{1}{2}}{\dd-\frac{3}{2}}\\
    &-\frac{1}{\q{\dd-\frac{3}{2}}}&0&1\\
    &&-\qfrac{2}{\dd-\frac{3}{2}}&0&1\\
    &&&&&1
  \end{pmatrix}
\end{equation}

\noindent
$V_\frac{1}{2}(\delta_1)\otimes V_\frac{3}{2}(\delta_2)$:
\begin{equation}
  F_{12}^{-1}=
  \begin{pmatrix}
    1\\
    &\qfrac{\dd-1}{\dd-2}\\
    &0&\qfrac{\dd}{\dd-2}\\
    &0&0&\qfrac{\dd+1}{\dd-2}\\
    &-\frac{1}{\q{\dd-2}}&0&0&1\\
    &&-\qfrac{2}{\dd-2}&0&0&1\\
    &&&-\qfrac{3}{\dd-2}&0&0&1\\
    &&&&&&&1
  \end{pmatrix}
\end{equation}

\noindent
$V_\frac{1}{2}(\delta_1)\otimes V_2(\delta_2)$:
\begin{equation}
  F_{12}^{-1}=
  \begin{pmatrix}
    1\\
    &\qfrac{\dd-\frac{3}{2}}{\dd-\frac{5}{2}}\\
    &0&\qfrac{\dd-\frac{1}{2}}{\dd-\frac{5}{2}}\\
    &0&0&\qfrac{\dd+\frac{1}{2}}{\dd-\frac{5}{2}}\\
    &0&0&0&\qfrac{\dd+\frac{3}{2}}{\dd-\frac{5}{2}}\\
    &-\frac{1}{\q{\dd-\frac{5}{2}}}&0&0&0&1\\
    &&-\qfrac{2}{\dd-\frac{5}{2}}&0&0&0&1\\
    &&&-\qfrac{3}{\dd-\frac{5}{2}}&0&0&0&1\\
    &&&&-\qfrac{4}{\dd-\frac{5}{2}}&0&0&0&1\\
    &&&&&&&&&1\\
  \end{pmatrix}
\end{equation}

\noindent
$V_1(\delta_1)\otimes V_1(\delta_2)$:
\begin{equation}
  F_{12}^{-1}=
  \begin{pmatrix}
    1\\
    &\qfrac{\dd}{\dd-2}\\
    &0&\frac{\q{\dd}\q{\dd+1}}{\q{\dd-2}\q{\dd-1}}\\
    &-\qfrac{2}{\dd-2}&0&1\\
    &&-\frac{\q{2}^2\q{\dd}}{\q{\dd-2}\q{\dd-1}}&0&\qfrac{\dd-1}{\dd-2}\\
    &&0&0&0&\qfrac{\dd}{\dd-2}\\
    &&\frac{\q{2}}{\q{\dd-2}\q{\dd-1}}&0&-\frac{1}{\q{\dd-2}}&0&1\\
    &&&&&-\qfrac{2}{\dd-2}&0&1\\
    &&&&&&&&1
  \end{pmatrix}
\end{equation}

\noindent
$V_1(\delta_1)\otimes V_\frac{3}{2}(\delta_2)$: The only non-vanishing
matrix elements are
\begin{mathletters}
\begin{eqnarray}
  F_{12}^{-1}\kk{0}{0} 
    &=& \kk{0}{0},\\
  F_{12}^{-1}\kk{0}{1} 
    &=& \qfrac{\dd-\frac{1}{2}}{\dd-\frac{5}{2}}\kk{0}{1} 
       -\qfrac{2}{\dd-\frac{5}{2}}\kk{1}{0},\\
  F_{12}^{-1}\kk{0}{2} 
    &=& \sfrac{\q{\dd-\frac{1}{2}}\q{\dd+\frac{1}{2}}}{\q{\dd-\frac{5}{2}}\q{\dd-\frac{3}{2}}}\kk{0}{2}
       -\sfrac{\q{2}^2\q{\dd-\frac{1}{2}}}{\q{\dd-\frac{5}{2}}\q{\dd-\frac{3}{2}}}\kk{1}{1}\nn\\
    &&+\sfrac{\q{2}}{\q{\dd-\frac{5}{2}}\q{\dd-\frac{3}{2}}}\kk{2}{0},\\
  F_{12}^{-1}\kk{0}{3}
    &=& \sfrac{\q{\dd+\frac{1}{2}}\q{\dd+\frac{3}{2}}}{\q{\dd-\frac{5}{2}}\q{\dd-\frac{3}{2}}}\kk{0}{3}
        -\sfrac{\q{2}\q{3}\q{\dd+\frac{1}{2}}}{\q{\dd-\frac{5}{2}}\q{\dd-\frac{3}{2}}}\kk{1}{2}\nn\\
    &&+\sfrac{\q{2}\q{3}}{\q{\dd-\frac{5}{2}}\q{\dd-\frac{3}{2}}}\kk{2}{1},\\
  F_{12}^{-1}\kk{1}{0}
    &=& \kk{1}{0},\\
  F_{12}^{-1}\kk{1}{1}
    &=& \qfrac{\dd-\frac{3}{2}}{\dd-\frac{5}{2}}\kk{1}{1}
        -\sfrac{1}{\q{\dd-\frac{5}{2}}}\kk{2}{0},\\
  F_{12}^{-1}\kk{1}{2}
    &=& \qfrac{\dd-\frac{1}{2}}{\dd-\frac{5}{2}}\kk{1}{2}
        -\qfrac{2}{\dd-\frac{5}{2}}\kk{2}{1},\\
  F_{12}^{-1}\kk{1}{3}
    &=& \qfrac{\dd+\frac{1}{2}}{\dd-\frac{5}{2}}\kk{1}{3}
        -\qfrac{3}{\dd-\frac{5}{2}}\kk{2}{2},\\
  F_{12}^{-1}\kk{2}{0}
    &=&\kk{2}{0},\\
  F_{12}^{-1}\kk{2}{1}
    &=&\kk{2}{1},\\
  F_{12}^{-1}\kk{2}{2}
    &=&\kk{2}{2},\\
  F_{12}^{-1}\kk{2}{3}
    &=&\kk{2}{3}.
\end{eqnarray}%
\end{mathletters}%

\noindent
$V_\frac{3}{2}(\delta_1)\otimes V_\frac{3}{2}(\delta_2)$: The only
non-vanishing matrix elements are
\begin{mathletters}
\begin{eqnarray}
  F_{12}^{-1}\kk{0}{0}
    &=& \kk{0}{0},\\
  F_{12}^{-1}\kk{0}{1}
    &=& \qfrac{\dd}{\dd-3}\kk{0}{1}
       -\qfrac{3}{\dd-3}\kk{1}{0},\\
  F_{12}^{-1}\kk{0}{2}
    &=& \sfrac{\q{\dd}\q{\dd+1}}{\q{\dd-3}\q{\dd-2}}\kk{0}{2}
        -\sfrac{\q{2}\q{3}\q{\dd}}{\q{\dd-3}\q{\dd-2}}\kk{1}{1}\nn\\
    &&  +\sfrac{\q{2}\q{3}}{\q{\dd-3}\q{\dd-2}}\kk{2}{0},\\
  F_{12}^{-1}\kk{0}{3}
    &=& \sfrac{\q{\dd}\q{\dd+1}\q{\dd+2}}{\q{\dd-3}\q{\dd-2}\q{\dd-1}}\kk{0}{3}\nn\\
    && -\sfrac{\q{3}^2\q{\dd}\q{\dd+1}}{\q{\dd-3}\q{\dd-2}\q{\dd-1}}\kk{1}{2}\nn\\
    && +\sfrac{\q{3}^2\q{2}\q{\dd}}{\q{\dd-3}\q{\dd-2}\q{\dd-1}}\kk{2}{1}\nn\\
    && -\sfrac{\q{2}\q{3}}{\q{\dd-3}\q{\dd-2}\q{\dd-2}}\kk{3}{0},
\end{eqnarray}

\begin{eqnarray}
  F_{12}^{-1}\kk{1}{0}
    &=& \kk{1}{0},\\
  F_{12}^{-1}\kk{1}{1}
    &=& \qfrac{\dd-1}{\dd-3}\kk{1}{1}
       -\qfrac{2}{\dd-3}\kk{2}{0},\\
  F_{12}^{-1}\kk{1}{2}
    &=& \sfrac{\q{\dd-1}\q{\dd}}{\q{\dd-3}\q{\dd-2}}\kk{1}{2}
        -\sfrac{\q{2}^2\q{\dd-1}}{\q{\dd-3}\q{\dd-2}}\kk{2}{1}\nn\\
    &&  +\sfrac{\q{2}}{\q{\dd-3}\q{\dd-2}}\kk{3}{0},\\
  F_{12}^{-1}\kk{1}{3}
    &=& \sfrac{\q{\dd}\q{\dd+1}}{\q{\dd-3}\q{\dd-2}}\kk{1}{3}
        -\sfrac{\q{2}\q{3}\q{\dd}}{\q{\dd-3}\q{\dd-2}}\kk{2}{2}\nn\\
    &&  +\sfrac{\q{2}\q{3}}{\q{\dd-3}\q{\dd-2}}\kk{3}{1},\\
  F_{12}^{-1}\kk{2}{0}
    &=& \kk{2}{0},\\
  F_{12}^{-1}\kk{2}{1}
    &=& \qfrac{\dd-2}{\dd-3}\kk{2}{1}
        -\sfrac{1}{\q{\dd-3}}\kk{3}{0},\\
  F_{12}^{-1}\kk{2}{2}
    &=& \qfrac{\dd-1}{\dd-3}\kk{2}{2}
        -\qfrac{2}{\dd-3}\kk{3}{1},\\
  F_{12}^{-1}\kk{2}{3}
    &=& \qfrac{\dd}{\dd-3}\kk{2}{3}
        -\qfrac{3}{\dd-3}\kk{3}{2},\\
  F_{12}^{-1}\kk{3}{0}
    &=& \kk{3}{0},\qquad
  F_{12}^{-1}\kk{3}{1}
    = \kk{3}{1},\\
  F_{12}^{-1}\kk{3}{2}
    &=& \kk{3}{2},\qquad
  F_{12}^{-1}\kk{3}{3}
    = \kk{3}{3}.
\end{eqnarray}%
\end{mathletters}

\acknowledgements

The author would like to thank DAAD for his scholarship. Thanks are
also due to A.~J.~Macfarlane, R.~Oeckl and F.~Wagner for valuable
discussions and comments.


\end{document}